# *Determining the Optimal Phase-Change Material via High-Throughput Calculations*

Nicholas A. Pike[*, 1], Amina Matt[2], and Ole M. Løvvik[1, 3]

[1]Center for Materials Science and Nanotechnology, University of Oslo, NO-0349 Oslo, Norway
[2]Institute of Materials, Swiss Federal Institute of Technology, Lausanne, Switzerland
[3]SINTEF Materials Science, Forskningsveien 1, NO-0314 Oslo, Norway
[*]Nicholas.pike@smn.uio.no

The discovery and optimization of phase-change and shape memory alloys remain a tedious and expensive process. Here a simple computational method is proposed to determine the ideal phase-change material for a given alloy composed of three elements. Using first-principles calculations, within a high-throughput framework, the ideal composition of a phase-change material between any two assumed phases can be determined. This ideal composition minimizes the interface strain during the structural transformation. Then one can target this ideal composition experimentally to produce compounds with low mechanical failure rates for a potentially wide variety of applications. Here we will provide evidence of the effectiveness of our calculations for a well-known phase-change material in which we predict the ideal composition and compare it to experimental results.

**INTRODUCTION:**

The discovery and understanding of phase-change materials is important due to the ever-growing presence of these materials in energy and medical applications. Phase-change materials can be exploited for energy storage [1, 2], medical devices [3], and as materials for energy conversion applications [4, 5] when magnetic [6], dielectric [7], or transport properties [8] of the material change due to the phase transition. Finding new systems is normally a time-consuming and expensive process. Not only do many samples need to be prepared and analyses, but one frequently needs to repeat this process several times to find the ideal composition with the preferred physical or mechanical property.

Experimentally, there have been many investigations into solid-to-solid state phase-change materials which undergo a martensitic phase transformation [9, 10, 11, 12] and therefore a change in the crystal structure. These transformations are diffusionless and may be highly-reversible due to a high level of geometric compatibility between the phases [13, 14]. Theoretical and mathematical efforts [13, 15] to describe this geometric compatibility via a set of mathematical equations have led to a set of conditions known as the "cofactor conditions". These conditions can be used to determine the composition of an alloy with the lowest interface energy for a particular phase transformation.

Here, we present our recent work using a first-principles approach to efficiently screen and predict the structural, electronic, and dielectric properties of a wide range of alloys using density functional theory (DFT) and density functional perturbation theory (DFPT) calculations within the Vienna Ab-initio Simulation Package (VASP) [16, 17]. Indeed, high-throughput screening based on DFT calculations have recently been shown [18] to be a very efficient method for selecting candidate compounds before a more detailed experimental search is undertaken.

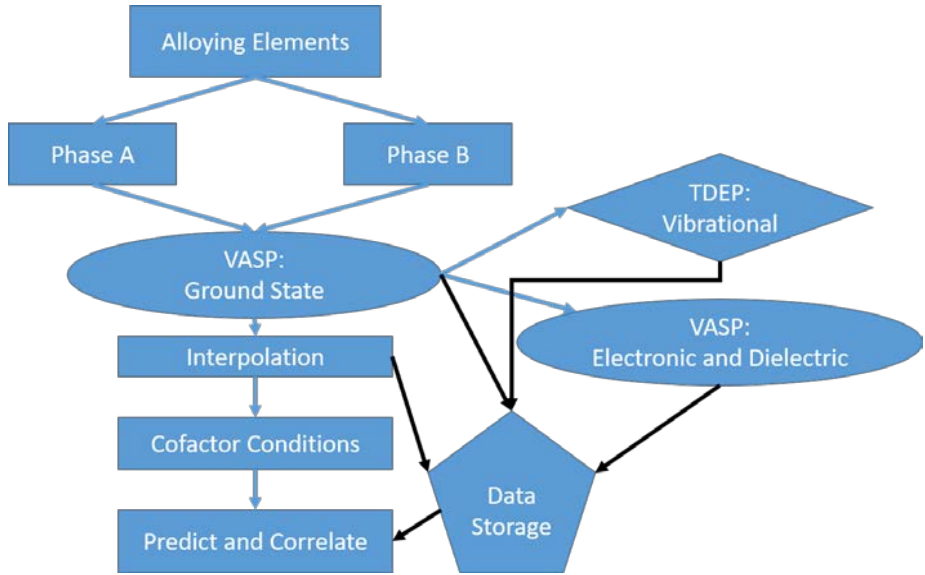

Figure 1: Workflow diagram of our high-throughput calculations of the structural, electronic, and dielectric properties of ternary alloys. Rectangles denote calculations done within our python-based code, ovals represent first-principles calculations with VASP, the diamond represents temperature dependent effective potential (TDEP) calculations and the pentagon corresponds to data storage solutions. Black lines refer to processes directly related to data storage and blue lines correspond to data transfer processes between different parts of the algorithm.

Utilizing the calculated structural properties of the compounds, Vegard's law, and the cofactor conditions, we are able to predict the composition of each alloyed compound with the lowest interface strain. Additionally, the information generated via our initial structural calculations leads to a wealth of data when we extended the calculations to include the electronic, dielectric, and vibrational properties of each calculated compound. While electronic and dielectric calculations are done using both DFT and DFPT within VASP, our calculations of the vibrational properties utilize the temperature dependent effective potential (TDEP) code [19, 20, 21]. The TDEP code extracts information from first-principles calculations to determine the temperature dependent vibrational properties of the system beyond the quasiharmonic approximation. These properties are stored in a database to help aid in the prediction of new materials and material properties via machine learning and data mining methods.

In what follows, we will first outline the methodology of our calculation by describing the algorithm used to generate the necessary first-principles calculations and extract information needed for later predictions of new phase-change materials. Next, we outline the results of this algorithm for a well-known phase change material, a ternary alloy composed of Au-Cu-Zn [11]. Finally, we will discuss future additions to our algorithm and the prospects of finding a new phase-change material and provide concluding remarks.

## METHODS

Figure 1 shows an outline of the workflow for our procedure where first-principles calculations using VASP are shown in ovals, rectangles represent steps processed via our python-based code, diamonds refer to post-processing calculations done using TDEP, and data storage solutions are indicated with a pentagon.

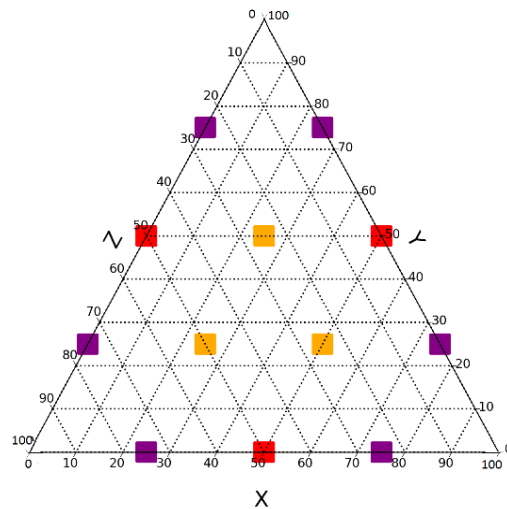

Figure 2: Ternary diagram highlighting the compositions with ideal ratios of atoms and the smallest unit cells. Yellow squares are Heusler (2:1:1) compounds, red squares are 1:1 compounds, and purple squares are 3:1 compounds.

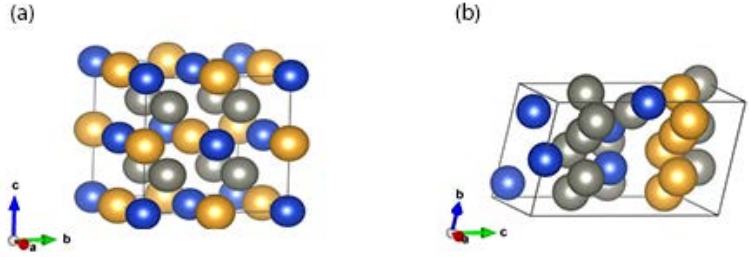

Figure 3: Crystal structures of Zn$_2$AuCu in (a) the face-centred L2$_1$ high-temperature phase and (b) the low-temperature 6M monoclinic phase.

Our process begins by defining a set of three elements and two crystal phases one wishes to investigate. With this information, our algorithm automatically generates the necessary VASP input files to undertake first-principles calculations via structural relaxation on a subset of compounds in the ternary phase diagram. As indicated in figure 2, the chosen twelve compounds are those with the smallest ratio of elements, and therefore the smallest unit cells, for a given phase. This minimizes the overall computational time necessary to calculate the structural information needed for the cofactor conditions, as shown below.

Relaxation calculations for all twelve compounds, in both crystal phases, occur within VASP. With the relaxed structural information, one can then launch additional calculations of each compound for the electronic and dielectric properties, as well as calculations of the vibrational properties via TDEP. Each of these first-principles or post-processing calculations can be done independently and in parallel allowing for an efficient use of supercomputing resources. During each stage of the calculation, our algorithm automatically stores practical information about each compound in each phase so that this information can be used to later predict and correlate materials properties and aid in the discovery of new phase-change materials.

**Calculation methods**

Our calculation of the structural properties uses DFT calculations within the VASP software package. The projector-augmented wave (PAW) method [22] was used in our calculations to describe the interaction between the ions and the valence electrons while utilizing PAW pseudopotentials [23]. Exchange and correlations effects were described by the Perdew-Burke-Ernzerhof (PBE) generalized gradient approximation (GGA) [24]. To describe the electronic system, a plane-wave energy cut-off equal to 500 eV was used and a Monkhorst-Pack [25] mesh of points was generated for each grid assuming a k-point density equal to five per Å$^{-1}$. For each ground state calculation, the iterations of the total energy were stopped once the differences in energy between successive iterations were less than 0.01 meV.

**Vegard's Law**

After our relaxation calculations, we extract the cell lattice parameters and interpolate them over the entire ternary phase diagram employing Vegard's law. Vegard's law, in its original form, states that for a binary phase alloy, the change in a physical property of a system should change linearly with respect to the atomic percent of one of the elements [26]. This means that, as the composition of a material changes smoothly from one compound to the next, one can linearly interpolate many of its physical properties. This can be extended to e.g. the lattice parameter a of the ternary alloy $A_{1-x}B_xC$ as

$$a[A_{1-x}B_xC] = (1-x)\,a[AC] + x a[BC], \qquad (1)$$

where AC and BC are binary compounds. For our purposes, we extend Eq. (1) to cover the entire compositional range, with the condition that the sum of the atomic fractions equals one. The functional form of our extended Vegard's law surface covers the entire compositional range of the ternary diagram and is given as

$$a(x,y) = x\alpha_1 + y\alpha_2 + (1 - x - y)\alpha_3 + \beta. \qquad (2)$$

In Eq. (2), the lattice parameter (a) is a function of the fractional composition (x) of element A, the fractional composition (y) of element B, and the fractional composition ($z = 1 - x - y$) of element C of an A-B-C ternary compound. Additionally, we define $\alpha_i$ for $i = 1,2,3$ as proportionality constants between the lattice parameter and composition and $\beta$ is a constant. During testing on the Au-Cu-Zn alloy below, our calculated lattice parameters fit Eq. (2) with $R^2 = 0.989$ for the cubic phase interpolation and gave an average value of $R^2 = 0.861$ for the monoclinic cell interpolation. However, not all compounds, especially Ag-Au and Ag-Cu based materials [27, 28], obey Vegard's law and therefore a quadratic extension of Eq. (1) may be necessary [29]. The need for a quadratic fit is indicated here for the monoclinic phase due to the value of $R^2$ for the linear interpolation. Therefore, our algorithm allows for an automatic extension of Eq. (2) to a quadratic order in composition.

With the interpolated lattice parameters and symmetry of the system, we can evaluate the cofactor conditions and determine the composition of the material that most closely satisfies the cofactor conditions. In principle, this requires only knowledge of the relaxed lattice parameters of the system and is therefore easily accessible with density functional theory.

## Cofactor Conditions

After gathering the structural parameters and symmetry of all twelve compounds, the cofactor conditions can be used to determine the composition with the lowest interface strain. The cofactor conditions were developed by Chen *et al.* [15] and, while their mathematical formulation is quite complex, the calculation of the cofactor conditions only requires the transformation matrix and the symmetries of the two phases. The geometric compatibility of the phases involved in a martensitic phase transformation is described in the crystallographic theory of martensitic transformations [30] as

$$\boldsymbol{F} = \boldsymbol{QU}. \qquad (3)$$

Here, **F** is the deformation matrix and **QU** is its polar decomposition where **Q** is a rotation matrix and **U** is the transformation stretch matrix. From our first-principles calculations the lattice parameters can be extracted, and the deformation matrix **F** can be determined [31]. In our case, we are particularly interested in determining the eigenvalues of the transformation stretch matrix **U**, which is easily derived from the matrix **F** as

$$\boldsymbol{U} = \sqrt{\boldsymbol{F}^T \boldsymbol{F}}. \qquad (4)$$

Note that only positive values of the square root are taken since $\boldsymbol{Q}^T \boldsymbol{Q} = 1$.

The middle eigenvalue of the transformation stretch matrix **U** corresponds to the first of the cofactor conditions, which are outlined in detail in Refs. [11, 15]. For our purposes, we wish to find the composition of the material such that the middle eigenvalue $\lambda_2 = 1$. The first cofactor condition is sufficient to ensure that there exists the possibility of a near-zero stress interface between the two phases of the material. Fulfilment of the other cofactor conditions, outlined in Ref. [11], ensure that the entire material transforms via super-compatibility of the phases.

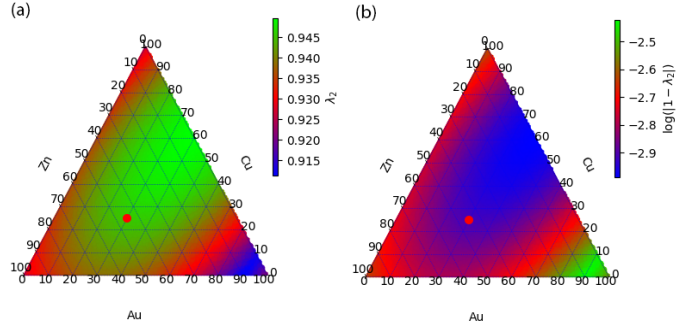

Figure 4: (a) Plot of the middle eigenvalue $\lambda_2$ vs composition for our Au-Cu-Zn ternary alloy system. (b) A plot of $\ln(|1 - \lambda_2|)$ vs composition. In each case, the literature value for the composition satisfying the cofactor conditions is given by a red dot at $Zn_{45}Au_{30}Cu_{25}$.

Using our algorithm and our first-principles calculations for the structural properties of twelve compounds shown in the figure 2, combined with our interpolation scheme based on Vegard's law. We can determine the composition of a ternary alloy that processes a middle eigenvalue closest to one. Next, we present our results on one such ternary alloy system, Au-Cu-Zn.

Table 1: Calculated lattice parameters and monoclinic angle for the twelve compounds shown in figure 2. Literature data for $Zn_2AuCu$ from Ref. [11] is listed, where the c lattice parameter is divided by three for comparison purposes and the monoclinic angle is modified to account for the reduced c lattice parameter.

|  | $a_0(Å)$ | $a(Å)$ | $b(Å)$ | $c(Å)$ | $\beta(°)$ |
|---|---|---|---|---|---|
| $AuCu$ | 6.211 | 4.806 | 5.528 | 13.993 | 76.819 |
| $AuZn$ | 6.378 | 5.098 | 5.532 | 14.243 | 77.579 |
| $CuZn$ | 5.939 | 4.633 | 5.230 | 13.244 | 77.157 |
| $Au_3Cu$ | 6.439 | 4.977 | 5.665 | 14.491 | 76.884 |
| $AuCu_3$ | 6.019 | 4.627 | 5.364 | 13.467 | 76.735 |
| $Au_3Zn$ | 6.514 | 4.835 | 6.012 | 14.764 | 74.091 |
| $AuZn_3$ | 6.297 | 4.684 | 5.657 | 13.809 | 76.875 |
| $Cu_3Zn$ | 5.852 | 4.518 | 5.185 | 13.162 | 76.570 |
| $CuZn_3$ | 6.047 | 4.701 | 5.437 | 13.280 | 76.478 |
| $Au_2CuZn$ | 6.289 | 4.940 | 5.532 | 13.856 | 76.735 |
| $Cu_2ZnAu$ | 6.094 | 4.690 | 5.392 | 13.760 | 76.160 |
| $Zn_2AuCu$ | 6.181 | 4.823 | 5.476 | 13.844 | 75.880 |
| $Zn_2AuCu$ [11] | 6.082 | 4.842 | 5.709 | 13.163 | 78.990 |

## RESULTS

Martensitic transformations in Au-Cu-Zn alloys have been known for many decades with initial investigations focused on increasing the transition temperature [32, 33]. While alloys of this material are known to be mechanically unstable and have poor thermomechanical performance [34] they have been shown to have exceptional mechanical properties when the composition is optimized [11]. For this ternary compound, the phase transition takes place between a high-temperature Heusler structure ($L2_1$) and a low–temperature modulated monoclinic structure known as *M18R* which is an orthorhombic phase (*Pnma*) if the shift in the atomic positions due to the modulation is ignored [11, 35, 36].

Therefore, in our first-principles calculations we have determined the structural parameters for all twelve compounds indicated in figure 2 in both the $L2_1$ cubic phase and a modulated monoclinic structure that is equal to 1/3 of the $M18R$ unit cell as shown by Otsuka et al. [37]. The optimized lattice parameters for both phases are given in Table 1 and compared to experimental measurements from Ref. [11].

With these lattice parameters and the interpolation techniques presented above we can find the deformation matrix **F** for each composition and find the middle eigenvalue of the transformation stretch tensor **U**. In figure 4, we display a plot of both the middle eigenvalue vs. composition and a plot of $\ln(|1 - \lambda_2|)$ vs. composition to highlight the compositions that satisfy the cofactor conditions.

Our preliminary results indicate the area in the phase diagram where the first cofactor condition is closest to being satisfied. In Ref. [11] it was found that a compound with the composition $Zn_{45}Au_{30}Cu_{25}$ nearly satisfied the cofactor conditions and indeed, our calculations indicate that this composition is within the area in figure 3 where the value of $\lambda_2$ is closest to one.

There are several possible reasons that we do not reproduce the experimental result precisely. The most important is probably that we did not use the proper ($M18R$) unit cell in our simulations. This is due to both the equivalence of the smaller cell as shown by Otsuka *et. al.* [37] and our desire for computational efficiency. With the unit cell that we did use, the lack of symmetry in the modulated structure makes first-principles calculations of some of the compounds difficult. For example, our first-principles calculations of $Au_3Zn$ resulted in lattice parameters which are considerably larger than other related compounds and leads to a possible source of error in our fitting procedure. As this is work in progress, we hope to see improvements in our correspondence with experimental data in other systems and through improvements in the construction of models and algorithms.

## CONCLUSIONS

We have calculated lattice parameters of the high-temperature cubic phase (austenite) and various low temperature phases (martensite) of an Au-Cu-Zu ternary alloy system. We find that the cubic phase is easily interpolated and leads to a high value of $R^2$. This allows us to have confidence in the lattice parameters for off-stoichiometric compositions. The monoclinic lattice constants show a quadratic trend and are interpolated with an extension of Vegard's law. In future work, we would like to confirm the accuracy of the lattice parameters for off-stoichiometric compositions using larger supercells to adequately represent the compositions of the off-stoichiometric compositions. Additionally, we would like to confirm our interpolated lattice parameters with experimental measurement of the lattice parameters of some off-stoichiometric compounds.

The method proposed here is a first step to define which compounds are susceptible candidates for satisfying the cofactor conditions and show a highly reversible martensitic transformation. Our calculations of the cofactor conditions for this single ternary alloy take considerably less time than an experimental search for the ideal compound. This speed and the information gained from the calculation of each of the twelve compounds in two different crystal phases are one of the defining features of this procedure. As more data is generated and the cofactor conditions are calculated for more compounds, we hope to develop connections between materials properties and shape memory properties that we can exploit in the future.

In particular, we wish to extend these calculations to other well-known phase-change materials to test the effectiveness of our calculations, but also to give us a starting point for future compounds with similar ternary alloys. It is our long-term goal to be able to predict new phase-change materials with tuneable and predictable materials characteristics. Additionally, we hope to develop an understanding of the mechanisms involved during the phase transformation by studying the vibrational properties of these materials throughout the phase-change process. It is the hope that this investigation will lead to a more thorough understanding of phase-change materials and the modulations observed in some of these compounds.

ACKNOWLEDGMENTS

We would like to thank the Research Council of Norway through the Frinatek program for funding. The computations were performed on resources provided by UNINETT Sigma2 the National Infrastructure for High-Performance Computing and Data Storage in Norway through grant numbers nn2615k and nn9462k.